\documentstyle[12pt]{article}

\newcommand{\eq}{\begin{equation}}
\newcommand{\en}{\end{equation}}
\newcommand{\bea}{\begin{eqnarray}}
\newcommand{\eea}{\end{eqnarray}}
\newcommand{\spz}{\hspace{0.7cm}}

\newcommand{\acc}{\\[3mm]}
\newcommand{\ba}{\begin{array}}
\newcommand{\ea}{\end{array}}
\newcommand{\ds}{\displaystyle}

\newcommand{\virg}{\spz,\spz}

\newcommand{\G}{{\cal G}}

\newcommand{\NP}[1]{Nucl.\ Phys.\ {\bf #1}}
\newcommand{\PL}[1]{Phys.\ Lett.\ {\bf #1}}

\newcommand{\MPL}[1]{Mod.\ Phys.\ Lett.\ {\bf #1}}
\newcommand{\IJMP}[1]{Int.\ J.\ Mod.\ Phys.\ {\bf #1}}

\begin{document}
\newcommand{\sps}{\hspace{3mm}}
\newcommand{\sts}{\footnotesize}
\newcommand{\stl}{\small}
\setlength{\unitlength}{1mm}
\newsavebox{\eotto}
\sbox{\eotto}{\begin{picture}(70,15)(0,-5)
\put(0,0){$E_8$ :}
\multiput(10,0)(10,0){7}{\circle*{1}}
\multiput(10,0)(10,0){6}{\line(1,0){10}}
\put(30,0){\line(0,1){10}}
\put(30,10){\circle*{1}}
\put(10,-1){\makebox(0,0)[t]{{\sts $2$}}}
\put(20,-1){\makebox(0,0)[t]{{\sts $6$}}}
\put(30,-1){\makebox(0,0)[t]{{\sts $8$}}}
\put(40,-1){\makebox(0,0)[t]{{\sts $7$}}}
\put(50,-1){\makebox(0,0)[t]{{\sts $5$}}}
\put(60,-1){\makebox(0,0)[t]{{\sts $3$}}}
\put(70,-1){\makebox(0,0)[t]{{\sts $1$}}}
\put(33,10){\makebox(0,0){{\sts $4$}}}
\end{picture}}

\renewcommand{\thefootnote}{\fnsymbol{footnote}}

\newpage
\setcounter{page}{0}
\begin{flushright}
Bologna preprint DFUB-92-12\\
Torino preprint DFTT-92-30\\
July 1992
\end{flushright}
\vskip .3cm

\vskip 1cm
\begin{center}
{\bf A NEW FAMILY OF DIAGONAL ADE--RELATED SCATTERING THEORIES}\\
\vskip 1.8cm
{\large F.\ Ravanini$^1$, R.\ Tateo$^2$ and A.\ Valleriani$^1$}
     \footnote{E-mail: ravanini@bologna.infn.it, tateo@torino.infn.it,
     valleriani@bologna.infn.it}\\
\vskip .7cm
{\em $^1$ I.N.F.N. - Sez. di Bologna, and Dip. di Fisica,\\
     Universit\`a di Bologna, Via Irnerio 46, I-40126 Bologna, Italy\\
\vskip .4cm
     $^2$ Dip. di Fisica Teorica, Universit\`a di Torino\\
     Via P.Giuria 1, I-10125 Torino, Italy}\\
\end{center}
\vskip 1cm

\renewcommand{\thefootnote}{\arabic{footnote}}
\setcounter{footnote}{0}

\begin{abstract}
\noindent
We propose the factorizable S-matrices of the massive excitations of the
non-unitary minimal model $M_{2,11}$ perturbed by the operator $\Phi_{1,4}$.
The massive excitations and the whole set of two particle S-matrices of the
theory is simply related to the $E_8$ unitary minimal scattering theory.
The counting argument and the Thermodynamic Bethe Ansatz (TBA)
are applied to this
scattering theory in order to support this interpretation. Generalizing this
result, we describe a new
family of {\em non unitary} and {\em diagonal}
$ADE$--related scattering theories. A further generalization suggests
the magnonic TBA for a large class of non-unitary $\G\otimes\G/\G$ coset models
($\G=A_{odd},D_n,E_{6,7,8}$) perturbed by $\Phi_{id,id,adj}$,
described by non-diagonal S-matrices.
\end{abstract}

\newpage

\section{Introduction}
As  pointed out by A.Zamolodchikov \cite{zam}, some deformations of
conformal field theories (CFT) retain an infinite number of integrals
of motion even away from criticality where the conformal invariance is
broken and a finite correlation length develops.
A.Zamolodchikov gave a sufficient condition, now known as the "counting
argument" to determine if some combinations of integrals of motion
of a given Lorentz spin survive the perturbation of the CFT.
If some do survive, and if the perturbed theory is purely massive,
then it can be described by a factorizable
S-matrix. The values of the  spins of the integrals of motion restrict
the possible bound state structure and mass ratios in the theory. The
bootstrap principle then allows one to actually conjecture a S-matrix
of the theory, and give  its particle content.
To check these conjectures for the S-matrices one can use the fact
that the infinite-volume
thermodynamics of a massive quantum field theory (QFT)
can be expressed  in terms of its
S-matrix. This method when applied to a factorizable S-matrix theory,
leads to the so called
Thermodynamic Bethe Ansatz (TBA) \cite{YY,Al-Potts}

In this letter we use the counting argument and the bootstrap procedure to find
the S-matrix of the $M_{2,11}$ minimal model perturbed by its relevant operator
$\Phi_{14}$. It turns out that this S-matrix is related to that of the Ising
model in a magnetic field and can be encoded on a $E_8$ Dynkin diagram.
Applying
the TBA technique, we confirm that this S-matrix actually describes $M_{2,
11}+\Phi_{14}$. The TBA turns out to be encoded on a kind of ``product'' of
graphs described in~\cite{wtba,RTV}, namely an $E_{8}$ diagram with all nodes
filled with tadpoles. At this point, the obvious generalization is to
consider the S-matrices related to TBA's encoded on any Dynkin diagram
decorated by tadpoles. The general tool to derive these S-matrices is a
powerful identity we recently found~\cite{RTV} for the class of $ADE$
S-matrices of~\cite{KM}. The problem to identify the models corresponding
to the newly introduced S-matrices is solved by applying again the TBA
method. The result is that these S-matrices describe some non-unitary minimal
models of $W$ algebras, perturbed by their field $\Phi_{id,adj}$. This fact
and the form of TBA (or better its encoding on pairs of graphs), suggests a
further generalization to entire series of non-unitary coset models of the
kind $\G\otimes \G/\G$.

\newcommand{\V}{{\bf V}}
\section{\bf Counting argument applied to
the $M_{2,11}+\Phi_{1,4}$ minimal non--unitary model}

It is possible to develop a method for finding the lowest
values of the conserved spins. This is based on the so called counting
argument.
Denote by $\hat{\V}_{s+1}$ the space of the quasi-primary descendents
of the identity operator at level $s+1$ i.e the factor space
\eq
\hat{\V}_{s+1}=\V_{s+1}/\partial_{z} \V_{s}
\label{c1}
\en
\newcommand{\W}{{\bf W}}
\noindent
Analogously, for the family $\W$ of the primary field $\Phi$ we consider
the factor space at level $s$.
\eq
\hat{\W}_{s+1}=\W_{s+1}/ \partial_{z} \W_{s}
\label{c2}
\en
\noindent
The mapping
\eq
\partial_{\bar{z}} : \hat{\V}_{s+1} \rightarrow g \hat{\W}_{s}
\label{c3}
\en
\noindent
has a non vanishing kernel if
\eq
\dim(\hat{\V}_{s+1}) > \dim( \hat{\W}_{s})
\label{c4}
\en
\noindent
In these circumstances there should exist
fields $T_{s+1}\in \hat{\V}_{s+1}$ and
$ \Phi_{s-1} \in \hat{\W}_{s-1}$ such that:
\eq
\partial_{\bar{z}} T_{s+1}=g \partial_{z} \Phi_{s-1}
\label{c5}
\en
\noindent
This imply for the theory with action
\eq
A=A^{*} +g \int \Phi(x) d^{2}x
\label{c6}
\en
\noindent
(where $A^*$ is the UV fixed point action) the existence of a
(local) conserved charge with spin s
\eq
P_{s}= \int (T_{s+1} dz+ g \Phi_{s-1} d\bar{z})
\label{c7}
\en
\noindent
The counting argument
for the $M_{2,11}$ non-unitary theory perturbed
by the operator $\Phi=\Phi_{1,4}$ (see table \ref{tab211})
shows the existence of a conserved current with spin $s=7$ but none
with spin $s=3,5$.
\begin{table}
\begin{center}
\begin{tabular}{||c|c|c|c|c|c|c|c|c|c||} \hline
 0 & -${ 4 \over 11}$ &
-${ 7\over11}$ &
-${ 9 \over 11}$ & -${ 10 \over 11}$ & -${ 10 \over 11}$ & -${ 9 \over 11}$
 & -${ 7\over11}$ & -${ 4 \over 11}$  & 0  \\ \hline
\end{tabular}
\parbox{10cm}{\caption{\protect{\footnotesize
                        Kac Table of the
                   conformal dimensions $\Delta_{pq}$ of the
                   $M_{2,11}$ model, where $p$ grows in vertical
         and $q$ in horizontal}}\label{tab211} }
\end{center}
\end{table}
The fusion rules
of this CFT do not have any internal symmetry. These two facts
together allow the possibility to
have the $" \phi^{3}"$ and the $"\phi^{2}_1 \phi_2 + \phi^{2}_2 \phi_1"$
property~\cite{zam}.
This fixes the mass ratio
${m_2 \over m_1}= {1+\sqrt{5}  \over 2} $ and the spin spectrum
$s=1,7,11,13,17,19,23,29~~ \bmod(30)$.
Now the experience with the bootstrap
suggest us to assume that there is another particle with mass ratio
${m_3 \over m_1} =2 \cos( {\pi \over 30})$
with the property $"\phi^{2}_1 \phi_3"$
and we propose for $S_{11}$ the form
\eq
S_{11}=F_{1\over15}(\theta)F_{1\over3}(\theta)F_{2\over5}(\theta)
       F_{-{1 \over 30}}(\theta)F_{-{11\over30}}(\theta)
\label{c8}
\en
\noindent
where
\eq
f_{x}(\theta)=
 {\sinh {1 \over 2} \left( \theta+ i\pi x \right) \over
  \sinh {1 \over 2} \left( \theta - i\pi x \right) }
\en
\eq
F_{x}(\theta)=f_{x}(\theta)f_{i\pi-x}(\theta)=
 {\tanh {1 \over 2} \left( \theta+ i\pi x \right) \over
  \tanh {1 \over 2} \left( \theta - i\pi x \right) }
\label{c9}
\en
\noindent
The remaining scattering amplitudes are obtained by induction,
applying the bootstrap \load{\footnotesize}{\sf} equation\footnote{A
simple pole of $S_{ab}$ at
$\theta_{ab}=i U^c_{ab}$ in the direct channel indicates that
there exists a bound state
$c$ of $a$ and $b$ whose mass is $m_c^2=m_a^2+m_b^2+2 m_a m_b cos(U^c_{ab})$}
\eq
S_{cd}(\theta)=S_{ad} (\theta+i \bar{U}^{b}_{ac})
S_{bd} (\theta-i \bar{U}^{a}_{bc})
\label{c10}
\en
\noindent
where $\bar{U}^{c}_{ab}=\pi-U^{c}_{ab}$. Finally the general
 S-matrix element $S_{ab}$ is given  by:
\eq
S_{ab}(\theta)={\tilde{S}_{ab} (\theta) \over
\prod_{k} \tilde{S}_{ak}^{G^{-1}_{bk} }(\theta)}
\label{c11}
\en
\noindent
where $\tilde{S}_{ab}$ is the S-matrix of the  Ising model with magnetic
field ( the theory $M_{3,4}+\Phi_{1,2}$) and,
$G^{-1}$ is the inverse of the incidence matrix of the Dynkin diagram of
$E_8$.

The exact mass spectrum is (as for $M_{3,4}+\Phi_{1,2}$):
\eq
\ba{lll}
m_1=m & m_2=2m\cos\frac{\pi}{5} &
m_3=2m\cos\frac{\pi}{30}\acc
m_4=2m_2\cos\frac{7\pi}{30} & m_5=2m_2\cos\frac{2\pi}{15} &
m_6=2m_2\cos\frac{\pi}{30}\acc
m_7=m_2\cos\frac{\pi}{5}\cos\frac{7\pi}{30} &
m_8=4m_2\cos\frac{\pi}{5}\cos\frac{2\pi}{15}. & \acc
\label{c12}
\ea
\en
for the correspondence between
the particles and the nodes in the Dynkin diagram
see fig.1.
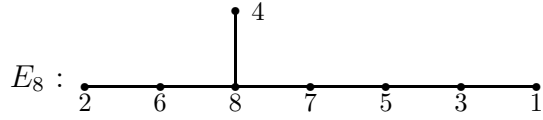
\begin{figure}
\begin{center}
\begin{picture}(70,40)(0,0)
\put(0,30){\usebox{\eotto}}
\put(0,20){\parbox{70mm}{\caption{\protect{\small
$E_8$ Dynkin diagram: the node labelled by $i$
corresponds to the mass $m_i$}}}}
\end{picture}
\end{center}
\label{fig1}
\end{figure}

\noindent
\section{\bf TBA for the $M_{2,11}+\Phi_{1,4}$ minimal non--unitary model}
The TBA equations, as they come from the
thermodynamic analysis of Bethe wave functions, have the following form:
\eq
-\nu_a(\theta)+\varepsilon_{a}(\theta)+{1 \over 2 \pi}
\sum_a [\phi_{ab}*\log(1+e^{-\varepsilon_{b}})](\theta)=0 \virg
 a=1,2,\cdots8
\label{c14}
\en
\noindent
where $\nu_{a}=R m_{a} \cosh(\theta)$ are the
energies (at rapidity $\theta$) of particles with the $E_8-$related mass
 spectrum $m_a$, and $\varepsilon_a(\theta)$ are
the so called  pseudoenergies,
corresponding to each species of particles in the spectrum.
The kernel $\phi_{ab}(\theta)$  in eq. (\ref{c14}),
encodes the scattering data thanks to its link with the S--matrix and
is given by
\eq
\phi_{ab}(\theta)= -i {d \over d \theta} \log S_{ab}(\theta)
\label{c16}
\en
\noindent
and the star $*$ stands for the rapidity convolution:
\eq
[\phi*L](\theta)= \int_{-\infty}^{+\infty} \phi (\theta-\beta)~ L(\beta) d\beta
\label{c17}
\en
\noindent
The pseudoenergies $\varepsilon_a(\theta)$ determine the Casimir energy
$E(R)={\pi \tilde{c}(R) \over 6R}$
 of the field theory on a circle of circumference $R$
\eq
\tilde{c}(R)={3 \over \pi^2} \sum_a
\int_{-\infty}^{+\infty} \nu_a(\theta)
\log(1+e^ {-\varepsilon_a(\theta)}) d\theta \virg
\tilde{c}=\tilde{c}(0)=c-24 \Delta_0
\label{c18}
\en
\noindent
where $c$ is the central charge and $\Delta_0$ the minimal conformal dimension
in the Kac-table. Now we use  the property introduced and proved in \cite{RTV}
for all $ADE$ S-matrices
\noindent
\eq
\tilde{S}_{ab} \left( \theta+i {\pi \over h} \right)
\tilde{S}_{ab} \left( \theta-i {\pi \over h} \right )=
e^{-i 2\pi \Theta(\theta) G_{ab}} \prod_c \tilde{S}_{ac}(\theta)^{G_{bc}}
\label{c119}
\en
where  the term proportional to the step function
\eq
\Theta(x)=\lim_{\epsilon\to 0}\left[\frac{1}{2}+\frac{1}{\pi}\arctan {x \over
\epsilon} \right] =
\left\{ \begin{array}{lll}
0 & {\rm if} & x<0 \\
\frac{1}{2} & {\rm if} & x=0 \\
1 & {\rm if} & x>0
\end{array} \right.
\en
has to be introduced to take into account the correct prescription for the
multivalued function $\log S$ and  $h=30$ is the Coxeter
number of $E_8$. Eq (\ref{c119}), together with (\ref{c11}) shows
 that $S$ must satisfy the equation
\noindent
\eq
S_{ab} \left( \theta+i {\pi \over h} \right)
S_{ab} \left( \theta-i {\pi \over h} \right )=
e^{-i 2\pi \Theta(\theta) (G_{ab}-\delta_{ab})} \prod_c S_{ac}(\theta)^{G_{bc}}
\label{c19}
\en
\noindent
The property (\ref{c19}), with the relation:
\eq
\sum_b G_{ab} m_b=2 \cos \left({\pi \over h} \right) m_a
\label{c20}
\en
\noindent
implies that the solutions to eqs  (\ref{c14}) are  also  particular solution
of the following functional "Y-system"
\eq
Y_a \left( \theta+{i \pi \over h} \right)
Y_a \left( \theta-{i \pi \over h} \right) =\prod_c (1+Y_c(\theta))^{G_{ac}}\
\left( 1+{1 \over Y_a(\theta)} \right) ^{-1}
\label{c21}
\en
\noindent
where $Y_a(\theta)=e^{\varepsilon_a(\theta)}$.
Eq (\ref{c21}) has the following periodicity:
\eq
Y_a \left( \theta+i \pi P \right)=Y_a(\theta) \virg P = { h+3 \over h}={11
\over
10}
\label{c22}
\en
\noindent
This can be shown (along the
lines of \cite{Zam1}) to be
in relation with the conformal dimension of the perturbing field
$\Phi$, via the formula
\eq
\Delta=1-{2 \over P}
\label{dDD}
\en
This allows to extract in a simple way the parameter $\Delta$,
characterizing, together with the effective central charge $\tilde{c}$,
the action of the theory.
The scaling function  energy
$F(R)={R E(R) \over 2
 \pi}$ expands in a regular series in $R^{\lambda}$
with $\lambda={4 h \over h+3}$
\eq
F(R)=- {\tilde{c} \over 12} +
{\epsilon_0 R^{2} \over 2\pi} + \sum_{k=0}^{\infty} f_k R^{\lambda k}
\label{c23}
\en
\noindent
where $\epsilon_0$ is the coefficient in the bulk energy term and can be
calculated from $S_{11}$ \cite{KM}
and
\eq
\tilde{c}=\sum_a  {6 \over \pi^2} L \left( {1 \over 1+ y_a} \right)
\label{c24}
\en
\noindent
where $L(x)$ is the Rogers dilogarithm function~\cite{Lewin}
\noindent
\eq
L=-{1 \over2 } \int_{0}^{+\infty} dy \left[{\log(y) \over 1-y} +{\log(1-y)
 \over y}\right]
\label{c26}
\en
\noindent
and $y_a$ are solutions of the equations
\eq
y_a^{2}=\prod_{c} (1+y_c)^{G_{ac}} {\left( 1+ {1 \over y_a} \right)}^{-1}
\label{c27}
\en
\noindent
Using the S-matrix elements (\ref{c11}) we find
\eq
\tilde{c}={8 \over 11}
\label{c28}
\en
\noindent
and from the equation (\ref{dDD})
\eq
\Delta= ~-{9 \over 11}
\label{c29}
\en
\noindent
i.e the correct values  of the effective central
charge of the theory $M_{2,11}$ and  the conformal
dimension  of the field $\Phi_{1,4}$ in  this theory.

\subsection{Universal form of TBA}
Now we  rewrite  eq. (\ref{c14}) in an appealing form which is
a generalization of the universal form~\cite{Zam1} of the TBA related
to the $ADE$ scattering theory. For our purposes we use the
identity  (\ref{c19}) and take its logarithmic derivative.
We find
\eq
\phi_{ab} \left( \theta+i {\pi \over h} \right) +
\phi_{ab} \left( \theta-i {\pi \over h} \right) =
\sum_c G_{bc} \phi_{ac}(\theta) -2 \pi \delta(\theta) (G_{ab}-\delta_{ab})
\en
Performing a Fourier transform this equation ($k$ is the momentum
 corresponding to $\theta$) becomes
\eq
2 \cos \left( {k \pi \over h} \right) \tilde{\phi}_{ab}(k)
= \sum_c G_{bc} \tilde{\phi}_{ac}(k) -2 \pi (G_{ab}-\delta_{ab})
\en
that allows us to rewrite the standard TBA system
in the form
\eq
\nu_a(\theta)=\varepsilon_a(\theta)+ {1 \over 2 \pi} \left[
 \sum_b G_{ab} \varphi_h *
(\nu_b-\log(1+e^{\varepsilon_b})) + \varphi_h * \log(1+e^{-\varepsilon_a})
\right](\theta)
\label{UN}
\en
where $\varphi_h(\theta)=\frac{h}{2\cosh \frac{h\theta}{2}}$ is the universal
kernel depending only on the Coxeter number $h_{E_8}=30$~\cite{Zam1}.
We recognize in this form the TBA denoted as $E_8 \diamond T_1$
in~\cite{RTV}\footnote{Along the lines of~\cite{RTV}
we denote  the "tadpole" diagram
$A_{2n}/Z_2$ as $T_n$, and we refer to that paper for a detailed explanation
of the symbol $G \diamond H$. }

\section{ADET-generalization}

The form of eq. (\ref{c11}) suggests a generalization to
all that cases where $G$ is the incidence matrix of some $ADET$
diagram\footnote{In the following we will refer to these models as
$ADET \diamond T_1$}.
In the following we  give a check of the validity of this hypothesis.
First of all it is necessary to
generalize eq.(\ref{c11})  to the cases where the incidence matrix
is not invertible, then we
compute the effective central charge for all the models in this class
and when $\tilde{c}$ corresponds to a minimal theory  we proceed to compare
our results with it.

Eq.(\ref{c11}) can be easily transformed using (\ref{c119}) and (\ref{c19})
as
\eq
S_{ab}\left(\theta+i\frac{\pi}{h}\right)S_{ab}
\left(\theta-i\frac{\pi}{h}\right)=
\tilde{S}_{ab}^{-1}(\theta)\tilde{S}_{ab}\left(\theta+i\frac{\pi}{h}\right)
\tilde{S}_{ab}\left(\theta-i\frac{\pi}{h}\right)
e^{-2\pi\delta_{ab} \Theta(\theta)}
\label{doesnot}
\en
This equation does not contain explicitly the $G^{-1}$ term, so we
can use it both for invertible and non--invertible incidence matrix.
Eq.(\ref{doesnot}) has a simple solution
\eq
S_{ab}(\theta)=\tilde{S}_{ab}(\theta)\varphi_{ab}^{-1}(\theta)
\label{vardoes}
\en
where $\varphi$ satisfy the relation
\eq
\varphi_{ab}\left(\theta+i\frac{\pi}{h}\right)
\varphi_{ab}\left(\theta-i\frac{\pi}{h}\right)=\tilde{S}_{ab}(\theta)
e^{2\pi\delta_{ab} \Theta(\theta)}
\label{Svar}
\en
The exponential term with the $\Theta$ function means  that
$\varphi$  has a single pole in $\theta={i \pi \over h}$.
We know that any $ADET$ diagonal $S$ matrix can always
be put in the form~\cite{KM,PD,MU}
\eq
\tilde{S}_{ab}(\theta)=\prod_{\alpha\in{\cal{A}}}f_{\alpha-\frac{1}{h}}(\theta)
f_{\alpha+\frac{1}{h}}(\theta)\virg
\en
where ${\cal{A}}$ is a set of rational numbers with common denominator $h$.
So $\varphi$ (\ref{Svar}) is expressible in terms of such blocks
\eq
\varphi_{ab}(\theta)=\pm\prod_{\alpha\in{\cal{A}}}f_{\alpha}(\theta).
\label{fifi}
\en
The ambiguity in
the sign in (\ref{fifi}) in the following is fixed by imposing
the fermion-like statistic of the system. Of course, when $G$ is invertible
$\varphi$ must have the form:
\eq
\varphi_{ab}=\pm \prod_c (\tilde{S}_{ac})^{G^{-1}_{bc}}
\label{FAB}
\en
Using this prescription the TBA equation and the Y-system can be put in a
universal form like (\ref{UN}) and (\ref{c21}) where now $G$ and $h$
are respectively
the incidence matrix and the Coxeter number of the corresponding
$ADET$ theory.The standard TBA central charge calculation gives
\eq
\tilde{c}_{UV} = r \left( 1- \frac{h}{h+3} \right)
\label{tta}
\en
where $h$ and $r$ are the Coxeter number and the rank of $\G$
respectively.
The dimension of the perturbing operator can be easily deduced
from the periodicity of the Y-system:
\eq
Y_a \left( \theta+i \pi P \right)=Y_{\bar{a}}(\theta)
\virg P = { h+3 \over h}
\label{c44}
\en
\noindent
In (\ref{c44}) $\bar{a}$ denotes the antiparticle of $a$ and if
the diagram does not possess any $Z_2$ symmetry $\bar{a}=a$.

We note from \ref{tta} that the  models denoted as $T_1\diamond T_1$,
$A_1\diamond T_1$ , $T_2\diamond T_1$ and $E_8 \diamond T_1$ have
central charge less than 1,
respectively
$\tilde{c}={1 \over 2},{3 \over 5} ,{3 \over 4},{8 \over 11}$.
Apart from
the theory $E_8 \diamond T_1$ studied above, the case $T_1 \diamond
T_1$ is
the Ising model perturbed with the thermal operator $\Phi_{1,3}$ of dimension
$\Delta= {1 \over 2}$.
This is in agreement with the $S_{11}=-1$ matrix obtained
using eq  (\ref{FAB}) and with $\Delta$ obtained from the formula
$\Delta=1- {1 \over P}$. The next, less trivial model is $A_1 \diamond
T_1$.
The $S$-matrix for the $A_1$ theory
is $\tilde{S}_{11}=-1$ and so from (\ref{fifi}) and (\ref{Svar})
we find for the theory $A_1\diamond T_1$ :
\eq
S_{11}(\theta)=f_{-{1\over 2}}(\theta)
\en
and using eqs (\ref{dDD},\ref{c44}) we find $\Delta= {1 \over 5}$.
We identify this model with
the minimal non-unitary theory  $M_{3,5}+\Phi_{1,3}$
proposed  in \cite{sm1} and studied in \cite{CM}.

Finally we study the theory $T_2 \diamond T_1$. The theory
$T_2$ has two particles (it is the theory called $A_4^{(2)}$
in~\cite{KM})
\eq
m_1=m\virg
m_2=m{{1+\sqrt{5}} \over 2}
\en
and S-matrix elements:
\eq
\ba{c}
\ds{\tilde{S}_{11}(\theta)=F_{2 \over 5}(\theta)  }\acc
\ds{\tilde{S}_{12}(\theta)=F_{1 \over 5}(\theta) F_{3 \over 5}(\theta)}\acc
\ds{\tilde{S}_{22}(\theta)= {F_{4 \over 5}(\theta)}
                            F_{2 \over 5}(\theta)}^2
\ea
\en
so using (\ref{c11}) it is simple to obtain
\eq
\ba{c}
\ds{S_{11}(\theta)=F_{2 \over 5}(\theta) F_{-{1 \over 5}}(\theta) }\acc
\ds{S_{12}(\theta)=F_{{1 \over 5}}}(\theta)\acc
\ds{S_{22}(\theta)=F_{2 \over 5}}(\theta)
\ea
\en
and, thanks to (\ref{dDD}),
$\Delta= -{ 1 \over 4}$. We identify this model as $M_{3,8}+\Phi_{1,3}$.
The set of two particle S-matrices is exactly that proposed in \cite{sm1,CM}

The remaining $\G \diamond T_1$ theories, for $\G=ADE$ also have
diagonal S-matrix. For $h$ even, it
is not difficult to figure out which are the
perturbed CFT's corresponding to these S-matrices.
They correspond to some non-unitary minimal models for the
$W\G$ algebra, namely $M_{h+1,h+3}^{W\G}$.\footnote{We denote here the
non-unitary minimal models of $W\G$ by $M_{pq}^{W\G}$, where, as usual, $p$
and $q$ are two coprimes.}
The dimension of the
perturbing operator, as given by eq.(\ref{dDD}), identifies it
with the $\Phi_{id,adj}$ field. The problem with $h$
odd is that $h+1$ and $h+3$ are not coprime.

This suggests an even more general identification of {\em magnonic} TBA's
of the form denoted in~\cite{RTV} by $(\G \diamond T_k)_l$, with the
non-unitary coset models of effective central charge
\eq
\tilde{c}_{UV}=\frac{dl}{l+h}\left(1-\frac{l(l+h)}{p(p+2l)}\right)
\en
where $d=\dim \G$ and $p=h+1,h+2,...$. The UV limit is then described by
the models $\G_k\otimes \G_l/\G_{k+l}$ with $k=\frac{p}{2}-h$. For $l=1$
the series of UV models is the $W\G$-minimal $M_{p,p+2}^{W\G}$. The perturbing
operator turns out to be always $\Phi_{id,id,adj}$ as usual. Although
these models have non-diagonal S-matrices, their TBA, here conjectured,
can give information on the structure of the RG flow of the model
off-criticality.

To conclude, we have found a class of diagonal S-matrices related to
$ADE$ Lie algebras, definitely different from those studied
in~\cite{KM}, that describe the perturbation of certain minimal
models of $W$-algebras. Further, we conjecture about a possible
extension of the TBA related to this models, to include {\em magnonic}
structures and allow entire classes of non unitary theories to be
followed along their RG flows.
We would like to emphasize that all these results have
been possible thanks to the high level of sophistication the TBA
technique has recently reached, and this is undoubtedly due to its deep
and intriguing link
with the theory of simply-laced Lie algebras and their Dynkin diagrams
first stressed in~\cite{Zam1} and
pursued in~\cite{Al,wtba,RTV}. This
observation not only allows an elegant systematization of known
results, but also provides a powerful tool for the study of the still
mysterious structure of the RG space of actions in two dimensions.

\vskip .6cm
{\bf Acknowledgements} -- We are greatly indebted to F.Gliozzi for a lot of
very
useful discussions. In
particular R.T. would like to thank F.Gliozzi for the patient and competent
tutoring during this first part of PhD in Torino. We also are grateful to
 A.Koubek, G.Mussardo and I.Pesando for useful discussions and help.
R.T. thanks the Theory Group at Bologna University for the
kind hospitality during the final part of this work.

\end{document}